\documentclass[pra,twocolumn,titlepage,nofootinbib,amsmath,showpacs]{revtex4}%
\usepackage{graphicx}

\begin{document}

\title{Contribution of Light-by-Light Scattering
to Energy Levels of Light Muonic Atoms}

\author{S.\,G.\,Karshenboim}
\altaffiliation[Also: ]{Max-Planck-Institut f\"{u}r Quantenoptik,
85748 Garching, Germany}
\email{s.g.karshenboim@vniim.ru}
\author{E.\,Yu.\,Korzinin}
\author{V.\,G.\,Ivanov}
\altaffiliation[Also: ]{The Central Astronomical Observatory
of the Russian Academy of Sciences at Pulkovo, 196140 St.Petersburg, Russia}
\author{V.\,A.\,Shelyuto}
\affiliation{D.~I. Mendeleev
Institute for Metrology, St.Petersburg, 190005, Russia}

\begin{abstract}
The complete contribution of diagrams with
the light-by-light scattering to the Lamb shift is found
for muonic hydrogen, deuterium and helium ion.
The results are obtained in the static muon approximation and a
part of the paper is devoted to the verification of this
approximation and analysis of its uncertainty.
\end{abstract}

\maketitle

Studies of energy levels in muonic hydrogen have a long history,
but until recently there have been no successful precision
measurements on this atom.
Recently results on the Lamb shift in muonic hydrogen and deuterium
were obtained for the first time \cite{mu_lamb}, and similar measurements
for muonic helium are planned.

The Lamb shift in light muonic two-body atoms is a splitting of the levels
$2s$ and $2p$, which are degenerate within a nonrelativistic (NR)
treatment of the Coulomb problem (as well as, e.g., in common
hydrogen). Effects of the electron vacuum polarization in muonic atoms
break down the degeneration and lead to the splitting of order
$\alpha(Z\alpha)^2m_\mu$, which can still be obtained from NR
calculations. (The relativistic units, in which $\hbar=c=1$, $Z$ is
the nuclear charge and $\alpha$ is the fine structure constant, are
applied throughout the paper.)

There are also ``finer'' splittings and, in particular,
levels in light muonic atoms possess the fine and hyperfine
structure, which result from relativistic calculations and appear
to be substantially smaller than the Lamb shift
(in contrast to common hydrogen).

Specificity of light muonic atoms is that the characteristic
atomic momentum is $Z\alpha m_\mu$,  comparable to the mass of
electron ($\alpha m_\mu\simeq 1.5\, m_e$). At the same time,
atomic energies are much lower than $m_e$.

Therefore, one can consider corrections to energy with closed electron loops
as nonrelativistic. Effects of the vacuum polarization up to
NR contributions to the Lamb shift of order $\alpha^5 m_\mu$
were studied in a number of papers (see, e.g., \cite{vp31,vp32,vp31e,rc}).

While calculating $\alpha^5 m_\mu$ terms in the papers mentioned,
the atomic nucleus
was treated as a point-like one. Nuclear-finite-size effects in light muonic
atoms can be considered, if necessary, as an additional perturbation.

\begin{figure}[thbp]
\begin{center}
{\includegraphics[width=0.95\columnwidth]{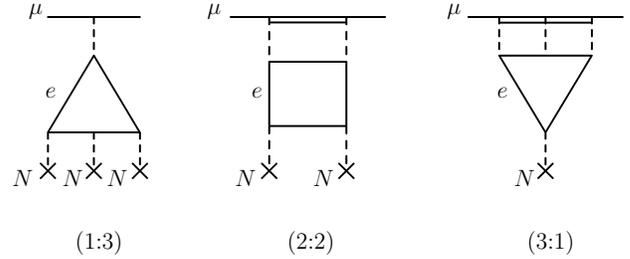}}
\end{center}
\caption{Three types of diagrams
for the LbL contributions to the Lamb shift in light muonic atoms
}
\label{fig:lbl}       
\end{figure}

Contributions, induced by the light-by-light (LbL) scattering
(see Fig.~\ref{fig:lbl}),
appear in order of  $\alpha^5 m_\mu$ and are considered in this paper.
The atomic nucleus in diagrams in Fig.~\ref{fig:lbl} is treated in the
external field approximation (i.e., as a static one),
while the expression for the muonic line includes
the Coulomb Green function of muon, which is indicated by
the double line,
\begin{equation}\label{gcoul}
S_C(E,{\bf p},{\bf p}^\prime)=i\sum_\lambda \frac{\vert\lambda({\bf
p})\rangle   \langle\lambda({\bf p}^\prime)\vert}{E-E_\lambda+i0}\;.
\end{equation}
The sum is taken over all intermediate states $\lambda$
of a discrete and continuous spectrum.

Each of the photons, including those that connect the muon line and
the electron
loop as well as those of the external nuclear field, are Coulomb
photons, i.e.,
only the $D_{00}$ component of the photon propagator contributes to
the result. $D_{00}$ does not depend on
energy in the Coulomb gauge, however, energy is transferred
through photons lines that connect the
muon line and the electron loop. Apparently, energy does not
propagate through the external field lines.

Generally speaking, since the muon atomic momentum
is of order of the electron loop momenta, explicit
forms of the functions $\vert\lambda({\bf p})\rangle$ are
substantially different for free and bound
(Coulomb) wave functions (see, e.g., \cite{EGSb}).

Calculations with Coulomb Green functions turn out to be rather
complicated even in the NR approximation; up to date they have not
been performed for any of the corresponding contributions.

\begin{figure}[thbp]
\begin{center}
{\includegraphics[width=0.99\columnwidth]{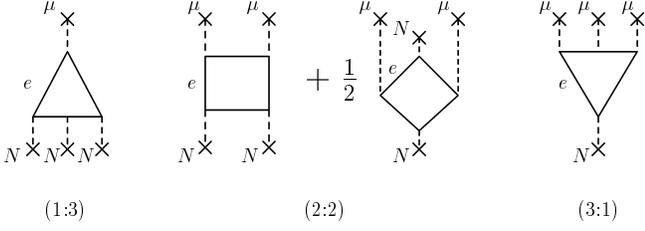}}
\end{center}
\caption{
Diagrams for the calculation of the Lamb shift contribution
in the static muon approximation
}
\label{fig:stat}       
\end{figure}

However, it is possible to demonstrate that the calculations can be made
in a simple approximation considering muon as a static particle
(see Fig.~\ref{fig:stat}).

In this approximation the expressions for individual contributions
to the energy of the level $\zeta$ are of the form
\begin{eqnarray}
\Delta E_{\rm 1:3}({\zeta})&=&(4\pi\alpha)\,(-4\pi Z\alpha)^3
\int\frac{d^3k_1}{(2\pi)^3}\int\frac{d^3k_2}{(2\pi)^3}\int\frac{d^3k_3}{(2\pi)^3}\nonumber\\
&&\times\frac1{{\bf k}_1^2\,{\bf k}_2^2\,{\bf k}_3^2\,{\bf k}_4^2}{\cal F}_{\zeta}({\bf k}_1)\cdot{\cal L} \;,\label{3:field13}\\
\Delta E_{\rm 2:2}({\zeta})&=&(4\pi\alpha)^2\,(-4\pi Z\alpha)^2
\int\frac{d^3k_1}{(2\pi)^3}\int\frac{d^3k_2}{(2\pi)^3}\int\frac{d^3k_3}{(2\pi)^3}\nonumber\\
&&\times\frac1{{\bf k}_1^2\,{\bf k}_2^2\,{\bf k}_3^2\,{\bf k}_4^2}\nonumber\\
&&\times\biggl[{\cal F}_{\zeta}({\bf k}_1+ {\bf k}_2) +\frac12 {\cal F}_({\bf k}_1+ {\bf k}_3) \biggl]\cdot {\cal L}\;,\label{3:field22}\\
\Delta E_{\rm 3:1}({\zeta})&=&(4\pi\alpha)^3\,(-4\pi Z\alpha)
\int\frac{d^3k_1}{(2\pi)^3}\int\frac{d^3k_2}{(2\pi)^3}\int\frac{d^3k_3}{(2\pi)^3}\nonumber\\
&&\times\frac1{{\bf k}_1^2\,{\bf k}_2^2\,{\bf k}_3^2\,{\bf
k}_4^2}{\cal F}_{\zeta}({\bf k}_4)\cdot {\cal L} \label{3:field31}
\;,
\end{eqnarray}
where
\begin{equation}\label{con:wv}
{\cal F}_{\zeta}({\bf
q})=\int\frac{d^3\!p}{(2\pi)^3}\;\Psi_{\zeta}^*({\bf
p})\;\Psi_{\zeta}({\bf p}+{\bf q})
\end{equation}
is the atomic form factor, which is for the $2p-2s$ splitting found to be
\begin{equation}\label{con:wv2ps}
{\cal F}_{2p-2s}({\bf q})=\frac{2\gamma^4{\bf q}^2(\gamma^2-{\bf
q}^2)}{(\gamma^2+{\bf q}^2)^4}\;,
\end{equation}
$\gamma=Z\alpha m_r$, and $m_r$ is the reduced mass of the muon in the atom.

\begin{figure}[thbp]
\begin{center}
{\includegraphics[width=0.17\columnwidth]{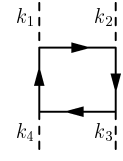}}
\end{center}
\caption{The diagram for the LbL scattering block
${\cal L}(k_1,k_2,k_3,k_4)$}
\label{fig:fbl}       
\end{figure}

The factor ${\cal L}(k_1,k_2,k_3,k_4)$ corresponds
to a single diagram of the LbL scattering with the momenta of the incoming
photons defined as $k_i=(0,{\bf k}_i)$ for $i=1,2,3$ and $k_4=-(k_1+k_2+k_3)$
(see Fig.~\ref{fig:fbl}). The permutations of photons,
as it is demonstrated below, have been already taken into account
explicitly.

We remark that both the Coulomb Green functions and
integrations over photon energy that are included in the full expressions
have completely vanished. Note that energy in the static approximation
does not propagate through photon lines, but does through
photons that connect the muon and electron.

The static muon approximation qualitatively changes the form
of expressions and notably simplifies the calculations.

We present here the proof of its applicability to light muonic atoms,
assuming at the first stage that the bound energy of the state
$\zeta$, for which the energy is calculated, as well as for intermediate
states $\lambda$, is negligible. Verification of this assumption
is discussed separately afterwards.

The standard muonic Coulomb Green function appears in our calculations
with the following arguments:
\[
S_C(E_\zeta\pm \omega,{\bf p},{\bf p}^\prime)\;,
\]
where $\omega$ is a photon frequency or combination of such values.
In contrast to the static muon approximation (see Fig.~\ref{fig:stat}),
these frequencies are not zeros in the full expression.

Neglecting the energies of the states and taking into account
completeness of the basis, one can find that the sum over
$\lambda$ becomes trivial:
\begin{eqnarray}\label{gcoul1}
\sum_\lambda \frac{\vert\lambda({\bf p})\rangle \langle\lambda({\bf
p}^\prime)\vert}{E_\zeta\pm \omega-E_\lambda+i0}&\to& \sum_\lambda
\frac{\vert\lambda({\bf p})\rangle   \langle\lambda({\bf
p}^\prime)\vert}{\pm\omega+i0}\nonumber\\
&=&\frac{(2\pi)^3\delta\bigl({\bf p}-{\bf p}^\prime\bigr)}{\pm
\omega+i0}\;.
\end{eqnarray}

Expressions with the Coulomb Green function differ from expressions
in the static muon approximation because three additional elements
appear:\\
1) a sum over intermediate states;\\
2) an integration over momentum (note, while a muon propagates at
the Coulomb field, its energy rather than momentum is conserved);\\
3) an integration over the photon frequencies (note that
the virtual photons, connecting the muon and electron in Fig.~\ref{fig:lbl},
transfer energy, while the photons in the static approximation
in Fig.~\ref{fig:stat} do not).

One should also take into account contributions of all
possible permutations of the photon lines. We treat in our calculations
a single diagram in Fig.~\ref{fig:fbl} as the whole LbL
scattering block, and all the permutations are taken into account
as contributions to the muonic factor with permuted lines of outgoing photons.

The transformation (\ref{gcoul1}) allows removing both
the sum over intermediate states and ``superfluous''
integrations over momentum. Now the only difference
between the expressions (\ref{3:field22}) and (\ref{3:field31})
and the Coulomb one is the integration over energy of the photons.
(The expression (\ref{3:field13}) for the 1:3-contribution
is already equal to the full expression since it does not really include
the Coulomb Green function from the beginning.)

\begin{figure}[thbp]
\begin{center}
{\includegraphics[width=0.7\columnwidth]{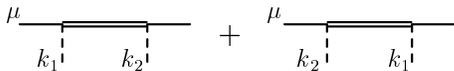}}
\end{center}
\vspace{-7pt} \caption{
The muon factor for the 2:2 contribution
}
\label{fig:f22}       
\end{figure}

Let us look at the structure of integrations over $\omega$.
Exchange of energy between the muon and electron
is present in the Coulomb diagrams, but complete energy transfer
from the nucleus to electron is equal to zero, and
therefore the sum of energies of all photons
connecting the electron and muon is zero as well.
Thus, in the case of the 2:2 contribution, having
defined the energy of the first photon $k_{10}$ as $\omega$
(so that $k_{20}=-\omega$), we find that the sum of
the direct and cross diagrams for the muon factor
(see Fig.~\ref{fig:f22}) leads to a $\delta$ function
\begin{equation}\label{g22}
\frac{i}{\omega +i0} + \frac{i}{-\omega +i0} =2\pi \delta(\omega)
\end{equation}
that removes the last ``superfluous'' integration. As a result,
we obtain the expression (\ref{3:field22}).

\begin{figure}[thbp]
\begin{center}
{\includegraphics[width=0.99\columnwidth]{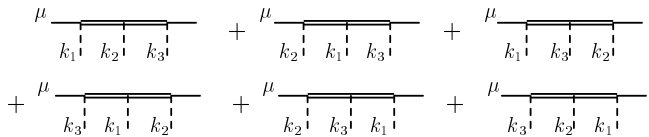}}
\end{center}
\caption{
The muon factor for the 3:1 contribution
}
\label{fig:f31}       
\end{figure}

The structure of the 3:1 diagrams is more complicated.
Having defined energies of the photons as
$k_{10}=\omega_1$, $k_{20}=\omega_2$
(so that $k_{30}=-(\omega_1+\omega_2)$), the sum of six permutations
for the total muon factor (see Fig.~\ref{fig:f31}) takes the form
\begin{eqnarray}
&&\frac{i}{\omega_1+i0}\,\frac{i}{\omega_1+\omega_2+i0}
+\frac{i}{\omega_2+i0}\,\frac{i}{\omega_1+\omega_2+i0}\nonumber\\
&&+\frac{i}{\omega_1+i0}\,\frac{i}{-\omega_2+i0}
+\frac{i}{-\omega_1-\omega_2+i0}\,\frac{i}{-\omega_2+i0}\nonumber\\
&&+\frac{i}{\omega_2+i0}\,\frac{i}{-\omega_1+i0} +
\frac{i}{-\omega_1-\omega_2+i0}\,\frac{i}{-\omega_1+i0}\nonumber\\
&=&\left(\frac{i}{\omega_1-i0}-\frac{i}{\omega_1+i0}\right)
\left(\frac{i}{\omega_2-i0}-\frac{i}{\omega_2+i0}\right)\nonumber\\
&=& 2\pi  \delta(\omega_1) \times 2\pi
\delta(\omega_2)\;.\label{delltaomega31}
\end{eqnarray}
The $\delta$ functions remove integrations over energy
and lead to the static expression (\ref{3:field31}).

Therefore, neglecting the energy of the states in the Coulomb
Green function (\ref{gcoul1}), we have in fact turned
the Coulomb diagrams in Fig.~\ref{fig:lbl} into diagrams
in Fig.~\ref{fig:stat} with a static muon.
We still have  to verify that it is really
possible to neglect the dismissed terms and
estimate the error of this procedure.

In case of the 2:2-contribution
the structure of the integral over the photon energy
for individual diagrams is of the form
\begin{equation}
{\cal I}(\epsilon)= \int {d\omega}\;\frac1{\pm\omega+\epsilon+i0}\;
{\cal L} (\omega)\;,
\end{equation}
where we define the combination of energies
that we are to neglect as $\epsilon\ll m_e$.

The first factor is for the muon  and it
is distinct for different contributions and the second one,
${\cal L} (\omega)$, which corresponds to the LbL scattering block,
is universal.

After combining with other diagrams the first factor
leads to the $\delta$ function at zero frequency (cf. (\ref{g22})
and (\ref{delltaomega31})), therefore, strictly speaking,
it cannot be expanded. One can see that after
performing a substitution of the variable by shifting
$\omega \to \omega^\prime\mp \epsilon$, the integral
turns to
\begin{equation}
{\cal I}(\epsilon)= \int
{d\omega^\prime}\;\frac1{\pm\omega^\prime+i0} \; {\cal L}
(\omega^\prime\pm \epsilon)\;.
\end{equation}
This expression can already be expanded in powers of $\epsilon$.

It is clear that while integrating in the LbL block,
the characteristic scale of loop energies in ${\cal L} (\omega)$ is
of order of the electron mass (or higher) and the parameter
of expansion is
\[
\frac{\epsilon}{m_e}\sim \frac{(Z\alpha)^2m_\mu}{m_e}\simeq 0.01\cdot Z^2\;.
\]

It is clear that a similar approach
with substitution of a variable is applicable for the 3:1 correction as well.
E.g., for the first term
in (\ref{delltaomega31}) one can write
\begin{eqnarray}
&&\int {d\omega_1}\int
{d\omega_2}\;\frac{1}{\omega_1+\epsilon_1+i0}\;\frac{{\cal L}
(\omega_1,\omega_2)}{\omega_1+\omega_2+\epsilon_2+i0}
\nonumber\\
&=&\int {d\omega^\prime_1}\int
{d\omega^\prime_2}\;\frac{1}{\omega^\prime_1+i0}\;\frac{{\cal L}
(\omega_1^\prime-\epsilon_1,\omega_2^\prime-\epsilon_2+\epsilon_1)}{\omega^\prime_1+\omega^\prime_2+i0}
\;,\nonumber
\end{eqnarray}
and then the expansion in $\epsilon$ can be done without problems.

We note that, if the diagrams in Fig.~\ref{fig:lbl} contain the
free muon propagators
instead of the Coulomb Green functions, the muon static approximation
is applicable and the
uncertainty is of the same order. Dealing with the the muon Green function (see, e.g., (\ref{gcoul1})), only the completeness
of the eigenstate basis is required, while the estimated eigenvalues of the energy are of the same order
for the free and Coulomb case, as long as the related integrals are convergent.

One of direct consequences of the applicability of the static muon
approximation and of symmetry of the expressions (\ref{3:field13})
and (\ref{3:field31}) is the identity \cite{rc}
\begin{equation}\label{3:1}
\Delta E_{3:1}=\frac{1}{Z^2}\Delta E_{1:3}\;.
\end{equation}

Until recently the 3:1 term has been the only completely unknown
contribution. On the contrary,
the 1:3-contribution, which is also referred to as the
Wichmann-Kroll contribution, has been well known
(see \cite{pach,EGS,EGSb} for results in muonic
hydrogen, \cite{bor_d} in muonic deuterium and \cite{bor_he,rc} muonic
helium). That is due to the fact that approximations
of the Wichmann-Kroll potential
in the form of an explicit function of distance, which are efficient
for calculations, have been known \cite{approx,rmp}.
There is also an exact
representation of this potential in the form of a double
integral \cite{blomquist},
understood in terms of the principal value.

The identity (\ref{3:1}) changes the situation radically. It by itself
improves accuracy of contributions to the Lamb shift
in muonic hydrogen and muonic deuterium by an order of magnitude and also leads to
a certain improvement of the accuracy for muonic helium.

According to Eq.~(\ref{3:1}), the 2:2 term, which is also referred to as
the virtual Delbr\"uck scattering, becomes the least accurately known
contribution from the LbL block. The results for muonic hydrogen
\cite{bor_h} and deuterium \cite{bor_d} are obtained with
accuracy less than 10\%, and for muonic helium-4 ion it has been rather
an estimate \cite{bor_he} than a result obtained.

The 2:2 contribution was examined in the cited papers in the so called
scattering approximation, where at first an operator related
to the Feynman diagram with external muon lines on the mass shell
and with a free muon propagator, is derived,
and then its matrix element over the Coulomb wave function
is calculated. In fact, before the calculation was really
done \cite{scattering,rmp},
a few additional approximations had been introduced, which reduced the
applied expressions to the static muon approximation.

Leaving aside the question about reasonability of using
the scattering approximation as an initial point,
we state that direct calculations
in \cite{bor_h,bor_d,bor_he} were in fact performed in the static muon
approximation, which is correct for the corresponding
Coulomb diagrams in Fig.~\ref{fig:lbl}
within the declared accuracy of calculations.

The 2:2 contribution is one of the most complex specific QED
contributions in light muonic atoms and absence of independent
confirmations significantly reduces reliability of the results
obtained in \cite{bor_h,bor_d,bor_he}.

Below we calculate the 1:3 and 2:2 terms. The first
calculation is used for control of analytical expressions
and numerical algorithms. The second one is aimed
to improve accuracy of the total contribution
of the diagram in Fig.~\ref{fig:lbl} for light muonic atoms,
since the accuracy of known results is not high enough.


To perform a calculation, one has at first to find an efficient form
of the factor ${\cal L}$,
which corresponds to the LbL scattering diagram (see Fig.~\ref{fig:fbl}).
A calculation within the standard technique of Feynman parameterization
leads to the result
\begin{eqnarray}
{\cal L}&=&{\cal L}(m_e)-{\cal L}(M)\,,\nonumber\\
{\cal L}(m)&=&\frac{1}{8\pi^2}\int_0^1{dx}\int_0^1{dy}\int_0^1{dz}\, x^2y \nonumber\\
&&\times \left[\frac{N_3}{2}\frac{m^4}{\Delta^2}
   -N_4\left(\frac{1}{\Delta}+\frac{m^2}{\Delta^2}\right)
   +\frac{N_6}{2}\frac{1}{\Delta^2}\right]\;,\label{13:3d}
\end{eqnarray}
where
\begin{eqnarray}
\Delta&=&  m^2-{\bf Q}^2+x\left({\bf k}_1^2+2y({\bf k}_1{\bf k}_2)\right.\nonumber\\
&&\left.+y{\bf k}_2^2+yz{\bf k}_3^2+2yz{\bf k}_3({\bf k}_1+{\bf k}_2)\right)\,,\nonumber%
\\
{\bf Q}& =&x\left( {\bf k}_1+y {\bf k}_2+yz {\bf
k}_3\right)\nonumber\,,\\
 N_3&=&4\,,\nonumber\\
N_4&=& -2\biggl[\left({\bf Q}-{\bf k}_2\right)\left({\bf Q}-{\bf
k}_1-{\bf k}_2-{\bf k}_3\right)\nonumber\\
&&+{\bf Q} \left({\bf Q}-{\bf k}_1\right)-{\bf k}_1 \left({\bf
Q}-{\bf k}_1-{\bf k}_2\right)\biggr]\,,\nonumber\\
N_6 &=& 4\biggl[\Bigl({\bf Q}\cdot({\bf Q}-{\bf k}_1)\Bigr)\Bigl(({\bf Q}-{\bf k}_1-{\bf k}_2)\nonumber\\
 &&~~~\cdot({\bf Q}-{\bf k}_1-{\bf k}_2-{\bf k}_3)\Bigr)\nonumber\\&&
-
 \Bigl({\bf Q}\cdot({\bf Q}-{\bf k}_1-{\bf k}_2)\Bigr)\nonumber\\&& ~~~ \times \Bigl(({\bf Q}-{\bf k}_1)\cdot({\bf Q}-{\bf k}_1-{\bf
k}_2-{\bf k}_3)\Bigr) \nonumber%
\\
&&+
 \Bigl({\bf Q}\cdot({\bf Q}-{\bf k}_1-{\bf k}_2-{\bf k}_3)\Bigr)\nonumber\\&&
~~~ \times \Bigl(({\bf Q}-{\bf k}_1)\cdot({\bf Q}-{\bf k}_1-{\bf
k}_2)\Bigr)\biggr] \,.\nonumber
 \end{eqnarray}

Here, we have explicitly introduced the Pauli-Villars regularization
with $M\gg m_e$, which can be useful if the convergence
of the integrals over $k$ is not good enough.

It is useful to make a further calculation by combining
the photon denominators and the denominator $\Delta$
using Feynman parameters. After integration over momenta
we come to the following expression:
\begin{equation}
\Delta E_{\rm 1:3}(\zeta)=\frac{3}{4\pi}\;\alpha(Z\alpha)^3
\int\frac{d^3q}{(2\pi)^3{\bf q}^2}{\cal F}_{\zeta}({\bf q})\;{\cal
J}_{\rm 1:3}\;,\label{13:uvw}
\end{equation}
where
\begin{eqnarray}
{\cal J}_{\rm 1:3}({\bf q}^2)&=& \int_0^1dx\int_0^1dy\int_0^1dz
  \int_0^1du\int_0^1dv\int_0^1dw\nonumber\\
&&\times
   \left\{{\cal A}_{1:3} \left[\ln\left(\frac{s_{1:3}\, {\bf q}^2+m_e^2}{m_e^2}\right)-\ln\left(\frac{M^2}{m_e^2}\right)\right]\right.
\nonumber\\
&&~~+   \left.
 \frac{{\cal B}_{1:3}\;{\bf q}^2}{\left(s_{1:3}\,{\bf q}^2+m_e^2\right)}
 +
\frac{{\cal C}_{1:3}\;{\bf q}^4}
 {\left(s_{1:3}\,{\bf q}^2+m_e^2\right)^2}
 \right\}\;,\label{1:3}
\end{eqnarray}
and the dimensionless coefficients ${\cal A}_{1:3}$, ${\cal B}_{1:3}$,
${\cal C}_{1:3}$ and $s_{1:3}$ are bulky functions of all Feynman
parameters.

Calculating the 1:3 contribution, the integration
over the momentum of the atomic form factor
${\cal F}$  in (\ref{3:field13}) is factorized,
and the remaining integrations over momenta involve logarithmic divergencies
at large momenta. Therefore, keeping $M$ at intermediate
stages turns to be useful for a calculation of individual terms.
In particular, contributions to (\ref{13:uvw}) of the separate terms of (\ref{13:3d})
include such divergences.

The value ${\cal J}_{\rm 1:3}({\bf q}^2)$ is nothing else but
a contribution to the charge form factor of the muon induced by
the LbL block. We can always ``renormalize''
the vertex function of muon and, by subtracting
\begin{equation}\label{renorm}
{\cal J}_{\rm 1:3}({\bf q}^2)\to {\cal J}_{\rm 1:3}({\bf q}^2) -
{\cal J}_{\rm 1:3}(0)\;,
\end{equation}
remove the logarithmic divergency in integration over $k$
in (\ref{1:3}).

In fact, the LbL scattering diagram, as it is known, does not
renormalize the vertex,
i.e., ${\cal J}_{\rm 1:3}(0)=0$, that we have checked both
numerically  and analytically
at different stages of transformations. In particular,
the coefficient at the logarithm $\ln(M/m_e)$ in (\ref{1:3})
turns to zero
\begin{eqnarray}
\int_0^1dx\int_0^1dy\int_0^1dz
  \int_0^1du\int_0^1dv\int_0^1dw\;
   {\cal A}_{1:3}
&=&0\nonumber\;.
\end{eqnarray}
Therefore ``renormalization'' of individual terms
according to (\ref{renorm}) is actually reduced
to regrouping and cancelation of individual divergent terms.

The final integration over Feynman parameters
and momentum $q$ has been performed by numerical integration
with VEGAS \cite{vegas}.
The results are presented in Table~\ref{t:13}.
The contributions for the $2p-2s$ splitting in the last line
of the table were calculated directly rather than as the difference
of the $2p$ and $2s$ terms.

 \begin{table}[htbp]
 \begin{center}
\begin{tabular}{cccc}
 \hline
Level(s)& $C(\mu{\rm H})$& $C(\mu{\rm
D})$ & $C(\mu{\rm He})$\\
 \hline
$2s$    & $~~0.5704(5)~$& $~~0.6263(5)~$ & $~1.0815(10)$  \\
$2p$    & $~~0.10468(3)$& $~~0.12405(3)$ &$~~0.5108(7)$ \\
$2p-2s$ & $ -0.4649(5)~$& $ -0.5015(5)~$ & $-0.5702(10)$ \\
 \hline
 \end{tabular}
\caption{
The 1:3 contribution (the Wichmann-Kroll contribution)
to energy levels of muonic hydrogen, deuterium and helium-4 ion:
$\Delta E=\alpha(Z\alpha)^4m_\mu \cdot10^{-3} \cdot C$
\label{t:13}}
 \end{center}
 \end{table}

The results for the Lamb shift $2p-2s$ (see also
Table~\ref{t:sum}) are in excellent agreement with the
values previously obtained by other authors
for muonic hydrogen ($-0.00103(2)\;$meV
\cite{EGS,EGSb}) and deuterium ($-0.00111(2)\;$meV \cite{bor_d}).
For helium ion our result agrees with $-0.0198(4)\;$meV \cite{rc}
and $-0.02\;$meV \cite{bor_he}, but
disagrees with the result 0.135\;meV \cite{mart}.
The uncertainty of the results \cite{EGS,bor_d} was not present
in the original papers
and was estimated here according to \cite{rc}.

The momentum integrations for the 2:2 contribution were performed
similarly to those described above.
One can check that it is possible here to set $M=\infty$
before integration over large momenta.
After introducing Feynman parameters and integrating over momenta, we obtain
\[
\Delta E_{\rm 2:2}(\zeta)=\frac{3}{4\pi}\;\alpha(Z\alpha)^3
\int\frac{d^3q}{(2\pi)^3}{\cal F}_{\zeta}({\bf q})\;{\cal J}_{\rm
2:2}\;,
\]
where
\begin{eqnarray}
{\cal J}_{\rm 2:2}({\bf q}^2)&=& \int_0^1dx\int_0^1dy\int_0^1dz
  \int_0^1du\int_0^1dv\int_0^1dw\int_0^1dt\nonumber\\
&&\times\sum_{k=1,2}
   \left\{
 \frac{{\cal B}^{(k)}_{2:2}}{\left(s_{2:2}^{(k)}\,{\bf q}^2+m_e^2\right)} +
\frac{{\cal C}^{(k)}_{2:2}\,{\bf q}^2}
 {\left(s^{(k)}_{2:2}\,{\bf q}^2+m_e^2\right)^2}
   \right.
\nonumber\\
&&+~~   \left. \frac{{\cal D}^{(k)}_{2:2}\,{\bf q}^4}
 {\left(s_{2:2}^{(k)}\,{\bf q}^2+m_e,^2\right)^3}
 \right\}\;.
\end{eqnarray}
The coefficients  ${\cal B}^{(k)}_{2:2}$, ${\cal C}^{(k)}_{2:2}$, ${\cal
D}^{(k)}_{2:2}$ and $s_{2:2}^{(k)}$ are bulky dimensionless functions
of the Feynman parameters. Note that the denominators for the ladder
($k=1$) and non-ladder ($k=2$) diagrams are slightly different.

The integration over $q$ and the Feynman parameters is performed
numerically. The results are presented in Table~\ref{t:22}.
Similarly to the case of the 1:3 contribution, the value for
the $2p-2s$ splitting in the table is found by a direct numerical integration.

 \begin{table}[htbp]
 \begin{center}
 \begin{tabular}{cccc}
 \hline
Level(s)& $C(\mu{\rm H})$& $C(\mu{\rm
D})$ & $C(\mu{\rm He})$\\
 \hline
$2s$     &  $-0.8201(15)$   & $-0.9000(15)$  & $-1.5615(25)$  \\
$2p$     &  $-0.2937(9)\ $  & $-0.3355(9)\ $ & $-0.908(2)~~~$ \\
$2p-2s$  & $~~0.5264(12)$   & $~~0.5645(12)$ & $0.652(2)~$  \\
\hline
 \end{tabular}
\caption{
The 2:2 contribution (the virtual Delbr\"uck scattering)
to energy levels of muonic hydrogen, deuterium
and helium-4 ion in the static muon approximation:
$\Delta E=\alpha^2(Z\alpha)^3m_\mu \cdot10^{-3} \cdot C$\label{t:22}}
 \end{center}
 \end{table}

The results for $2p-2s$ in the static muon approximation
are slightly smaller and have substantially higher accuracy
than the results
of the other authors. Ours are mostly in agreement with the former
values. In particular, for muonic hydrogen we obtain
0.001\,151(4)\;meV (cf. 0.001\,35(15)\;meV \cite{bor_h}),
for deuterium, 0.001\,234(4)\;meV (cf. 0.001\,47(16)\;meV \cite{bor_d})
and for helium ion, 0.01140(4)\;meV (cf. 0.02\;meV \cite{bor_he}).

Complete contributions of the diagram in Fig.~\ref{fig:lbl}
to the Lamb shift are collected in Table~\ref{t:sum}. In contrast to
Table~\ref{t:22}, where the uncertainty of
numerical calculations in the static muon approximation is presented,
here we take into account the error of the static muon approximation,
which dominates.

 \begin{table}[htbp]
 \begin{center}
 \begin{tabular}{cccc}
 \hline
Term& $\Delta E(\mu{\rm H})$& $\Delta E(\mu{\rm
D})$ & $\Delta E(\mu{\rm He})$\\
 & [meV]& [meV] & [meV]\\
 \hline
1:3 & $-0.001\,018(4)$ & $-0.001\,098(4)$ & $-0.019\,95(6)$ \\
2:2 &  $~~0.001\,15(1)$& $~~0.001\,24(1)$ & $~~0.0114(4)$ \\
3:1 & $-0.001\,02(1)$ & $-0.001\,10(1)$ & $-0.0050(2)$ \\
\hline
Total  &$-$0.000\,89(2)& $-$0.000\,96(2)& $-$0.0136(6) \\
 \hline
 \end{tabular}
\caption{
Contribution of the LbL scattering effects (Fig.~\ref{fig:lbl})
to the Lamb shift ($2p-2s$) in muonic
hydrogen, deuterium and helium-4 ion
\label{t:sum}}
 \end{center}
 \end{table}

In conclusion, we note smallness of the numerical coefficients
in Tables~\ref{t:13} and~\ref{t:22}, which is typical
for spin-independent nonrelativistic contributions of the light-by-light
scattering to energy levels.

As we have noted, the results for the 2:2 contribution obtained
in this paper are slightly below the results of Borie
\cite{bor_h,bor_d,bor_he}. The difference for muonic hydrogen
and deuterium is about 20\%, i.e. less than 1.5 standard deviations.
In the case of muonic helium our result (0.01140(2)\;meV)
is also lower than that of Borie (0.02\;meV),
however, the latter is presented in \cite{bor_he} in such a form
that one is rather able to estimate the scale of uncertainty
than its magnitude.

This difference in the results can in no way be considered
as a contradiction, but it has a systematic character
(all our results are lower than those of Borie)
and, strictly speaking, requires some explanation.

Due to that, we have checked two most important
elements of our calculations, namely, the expression
for the LbL scattering (\ref{13:3d}) and the integration
over momenta $k_i$ in (\ref{3:field31}). As a test
of the representation for $\cal L$ we have used a calculation
of the well-known 1:3 contribution (see Table~\ref{t:13})
and studied the expression (\ref{13:3d}) for the LbL scattering
in the case when momentum of one of the photons is zero.

To be certain that there are no systematic errors
in integration over momenta after introducing Feynman parameters
we have made an independent calculation of the 2:2 contribution,
in which integrations over momenta $k_i$ in the corresponding expression
(\ref{3:field31}) were performed directly. The results turn out to
have accuracy 4\% for muonic hydrogen and 3\% for muonic helium ion
and are in excellent agreement with results of Table~\ref{t:22}.

These tests allow to consider our results to be highly reliable.

The work was supported by DFG (Grant No. GZ 436 RUS 113/769/0-3).
The work of E.Y.K. was also supported by the Dynasty Foundation.

\end{document}